# Tilted Beam Measurement of VLBI Receiver for the South Pole Telescope

Junhan Kim[1,*] and Daniel P. Marrone[1]
[1]Department of Astronomy and Steward Observatory, University of Arizona, Tucson, AZ 85721, USA
*Contact: junhankim@email.arizona.edu

*Abstract*—We have developed a 230 and 345 GHz very-long-baseline interferometry (VLBI) receiver for the South Pole Telescope (SPT). With the receiver installed, the SPT has joined the global Event Horizon Telescope (EHT) array.
The receiver optics select the 230 or 345 GHz mixers by rotating the tertiary mirror around the optical axis, directing the chief ray from the secondary mirror to the feed horn of the selected frequency band. The tertiary is installed on top of the receiver cryostat, which contains both mixer assemblies. The feed horns are placed symmetrically across the centerline of the telescope optics and tilted inward by 5.7 degrees from the vertical plane so that their beams intersect at the chief ray intersection on the tertiary mirror.
We have performed vector beam measurements of the SPT VLBI receiver in both frequency bands. The measurements preserved the relative location of the beams, to establish the relative locations of the phase centers of the two horns. Measurements in two parallel reference planes above the cryostat were used to suppress reflected light. To model the beam, we derive a general expression of the electric field vector on the measurement plane for a tilted beam and infer the feed horn position parameters for both frequency bands by fitting models to data with a Markov chain Monte Carlo (MCMC) method. The inferred parameters such as the tilt angle of the feed horn are in good agreement with the design. We present the measurement setup, amplitude and phase pattern of the beam, and the fitting result here.

## I. Introduction

The South Pole Telescope (SPT) is a cosmic microwave background (CMB) experiment located at the Amundsen-Scott South Pole station [1,2]. Its 10-meter diameter primary dish was designed with a surface accuracy suitable for submillimeter-wavelength observation. Its geographic location, far from other submillimeter telescopes, provides an opportunity to greatly increase the size millimeter wavelength very-long-baseline interferometry (VLBI) arrays, such as the Event Horizon Telescope (EHT) [3].

We have developed a 230 and 345 GHz VLBI receiver for the SPT. In this paper, we report the vector beam measurement of the receiver using a technique developed for submillimeter receivers (e.g., [4,5]). We use the measurement to characterize the location and tilt of the horn phase centers inside the receiver dewar. In section 2, we describe the receiver and its optical design. In section 3, we show the beam measurement setup and the test result. In section 4, we introduce a general expression of the electric field when the tilt angle of the beam is considered, and then infer the model parameters using Markov chain Monte Carlo (MCMC) method. We have found that the estimated model parameters from the data agree well with the design.

## II. SPT VLBI Receiver

The SPT VLBI receiver operates in the 230 and 345 GHz frequency bands. Fig. 1 shows the inside of the dewar. In each band, the RF components – feed horns, ortho-mode transducers (OMT, a polarization separation device), and mixer blocks – are cooled down to 4 K. The 230 GHz receiver uses ALMA band 6 sideband-separating SIS mixers [6]. The 345 GHz mixer is under development and expected to be installed in early 2019.

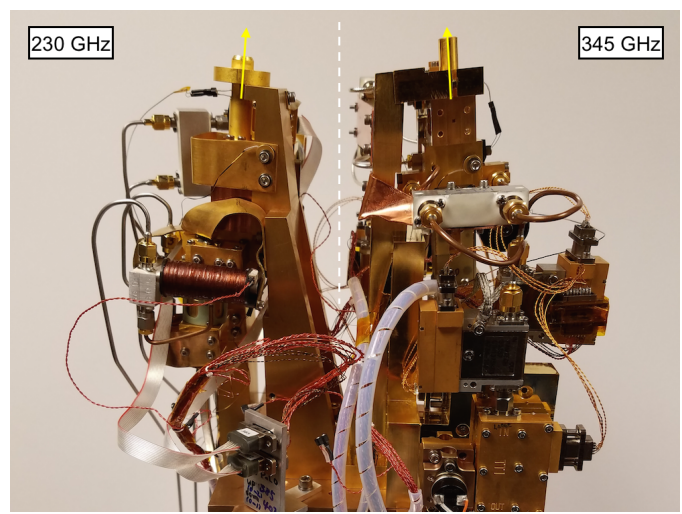

Fig. 1. An interior view of the SPT VLBI receiver. The left and right are 230 GHz and 345 GHz assemblies of the receiver, respectively. Both 230 GHz and 345 GHz feed horns are designed to be 5.74 degrees tilted inward so that the receiver operates at both frequencies using a rotating tertiary mirror above the dewar. The yellow arrows show the beam propagation directions of the feedhorns, and the white dashed line indicates the vertical plane.

The sky signal is delivered to the VLBI receiver with removable secondary and tertiary mirrors that are, independent of the SPT-3G optics (for details, see [7]). The tertiary mirror assembly sits on the receiver dewar and the mirror rotates around the optical axis to switch between 230 and 345 GHz. To couple the chief ray reflected through the tertiary mirror to each feed horn, we designed the feed horns to be tilted toward each other by 5.74 degrees.

### III. VECTOR BEAM MEASUREMENTS

Accurate positioning of the two feedhorns is critical to the coupling of the two frequency bands to the optics in this submillimeter-wavelength receiver. We perform 230 and 345 GHz vector beam measurement of the SPT VLBI receiver to characterize their location. We set up the vector beam measurement system as shown in Fig. 2. The test tone is generated from a Gunn oscillator followed by a frequency tripler (230 GHz: VDI WR-5.1 × 3, 345 GHz: VD1-WR2.8 × 3). We put WR-4 open waveguide probe (0.04 inch × 0.02 inch) whose cutoff frequency is ~140 GHz, on the tripler output, and the transmitter system is mounted on X-Y translation stage composed of two Parker Motion 403XE linear stages. The test tone frequency is chosen such that the intermediate frequency (IF) of the receiver is placed within the IF amplifier operating range. The IF is again mixed down to ~546 MHz to use a K&L Microwave bandpass filter with 10 MHz bandwidth to increase the signal to noise ratio. We can read the amplitude and relative phase of the second IF with the reference input from HP 8648C signal generator. During the beam scan, all the hardware components around the setup are covered with AN-72 broadband microwave absorber to reduce the reflection of the injected tone.

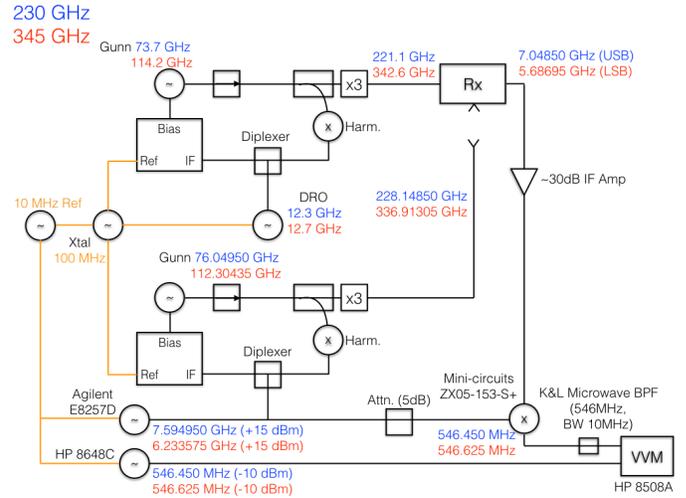

Fig. 2. Schematic of the beam measurement system. Frequencies for the 230 GHz chain are listed in blue, with 345 GHz in red. The signal generators and the phase locked loop for the system are locked to a common 10 MHz signal.

We perform a two-dimensional scan of a region 100 mm × 100 mm in size with a 2.5 mm step, in two parallel planes: The planes are ~200 mm and ~150 mm vertically above the horn phase center plane. The 230 GHz scan region is centered on $(x, y) = (25$ mm, $0$ mm$)$ and the 345 GHz region is centered on $(x, y) = (-25$ mm, $0$ mm$)$, to locate the maximum amplitude position near the center of the scanning area. After every column scan, the probe moves to the center position to track amplitude and phase stability. Interpolating reference measurements compensates the time-dependent phase drift during the scan. At each plane we make two maps on planes that are a quarter-wavelength apart in the $z$-direction. Averaging these two sets of data can further reduce the reflection. Fig. 3 shows the amplitude and phase of 230 GHz at $z = 200$ mm plane. For 230 GHz, we recorded the beams of two different polarizations, by rotating the waveguide probe by 90 degrees.

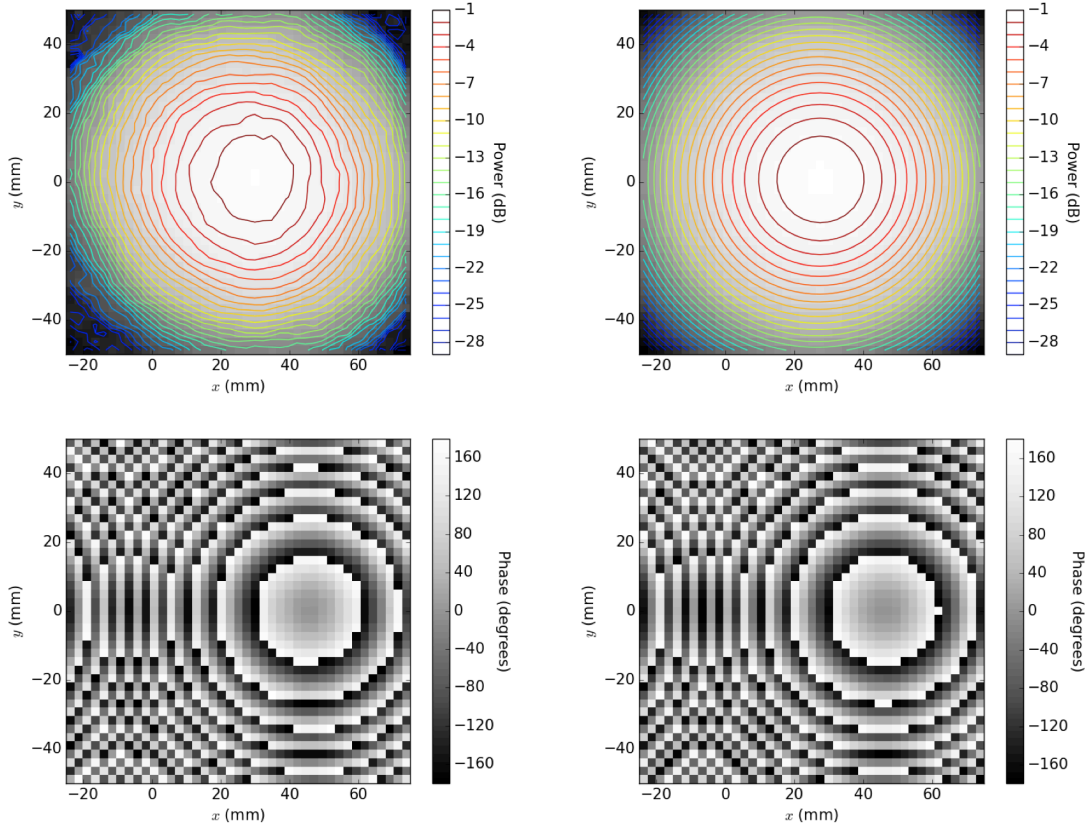

Fig. 3. (*Top left*) The amplitude and (*bottom left*) the phase pattern of 230 GHz beam measurement at $z = 200$ mm. The contours in the amplitude map are in dB units, and the phase map is plotted in degree units. The amplitude map shows more than 30 dB of dynamic range. The phase is not centered due to the tile of the feedhorn. (*Top right*) The amplitude and (*bottom right*) the phase pattern of the beam, using the analytical model with the best-fit parameters.

## IV. ANALYSIS

To analyze the data, we derive a general form of the electric field distribution of the tilted Gaussian beam, when the scanning plane is not vertical to the axis of propagation. Then, we fit the model to the data to infer the feed horn position and its tilt angle.

### A. Gaussian Beam Propagation of tilted horn

The normalized electric field distribution [8] is

$$E(r,z) = \left(\frac{2}{\pi w^2}\right)^{0.5} \exp\left(-\frac{r^2}{w^2} - jkz - \frac{j\pi r^2}{\lambda R} + j\phi_0\right),$$

where $z$ is the distance along the axis of propagation and $r$ is the perpendicular distance from the axis of propagation. $R$, $w$, and $\phi_0$ are the radius of curvature, beam radius, and the Gaussian phase shift and expressed as

$$R = z + \frac{1}{z}\left(\frac{\pi w_0^2}{\lambda}\right)^2,$$

$$w = w_0\left[1 + \left(\frac{\lambda z}{\pi w_0^2}\right)^2\right]^{0.5},$$

$$\phi_0 = \arctan\left(\frac{\lambda z}{\pi w_0^2}\right),$$

where $w_0$ is the beam waist radius.

In Fig. 4, we first define two planes parallel to each other: the horn phase center plane and the scanning plane where the waveguide probe mounted on the translation stage moves. We assume arbitrary horn phase center position $P_h = (x_h, y_h, 0)$ to derive an electric field at the scan position $P_s = (x, y, z_0)$.

The equation of the line $l_{beam}$ is

$$l_{beam}: (x_h, y_h, 0) + t(-\tan\theta_x, \tan\theta_y, 1).$$

The directional vector of axis of propagation on $l_{beam}$ is

$$\vec{b} = (-\tan\theta_x, \tan\theta_y, 1),$$

and the vector between $P_h$ and $P_s$ is

$$\overrightarrow{P_h P_s} = (x - x_h, y - y_h, z_0).$$

The vertical distance from the horn phase center to the wave plane that hits $P_s$, $z_{tilt}$ is

$$z_{tilt} = \frac{|\vec{b} \cdot \overrightarrow{P_h P_s}|}{|\vec{b}|},$$

where

$$|\vec{b} \cdot \overrightarrow{P_h P_s}| = |z_0 - (x - x_h)\tan\theta_x + (y - y_h)\tan\theta_y|,$$

and

$$|\vec{b}| = \sqrt{\tan^2\theta_x + \tan^2\theta_y + 1}.$$

The offset from the axis of propagation on the reference plane is

$$r_{tilt} = \frac{|\vec{b} \times \overrightarrow{P_h P_s}|}{|\vec{b}|},$$

where

$$\begin{aligned}|\vec{b} \times \overrightarrow{P_h P_s}|^2 &= [z_0 \tan\theta_y - (y - y_h)]^2 \\&+ [z_0 \tan\theta_x + (x - x_h)]^2 \\&+ [(y - y_h)\tan\theta_x - (x - x_h)\tan\theta_y]^2 \\&= z_0^2(\tan^2\theta_x + \tan^2\theta_y) \\&+ (x - x_h)^2(\tan^2\theta_y + 1) \\&+ (y - y_h)^2(\tan^2\theta_x + 1) \\&+ 2z_0(x - x_h)\tan\theta_x + 2z_0(y - y_h)\tan\theta_y \\&+ 2(x - x_h)(y - y_h)\tan\theta_x \tan\theta_y \\&= z_0^2(\sec^2\theta_x + \sec^2\theta_y) + (x - x_h)^2\sec^2\theta_y \\&+ (y - y_h)^2\sec^2\theta_x \\&- 2z_0[z_0 - (x - x_h)\tan\theta_x][z_0 \\&+ (y - y_h)\tan\theta_y].\end{aligned}$$

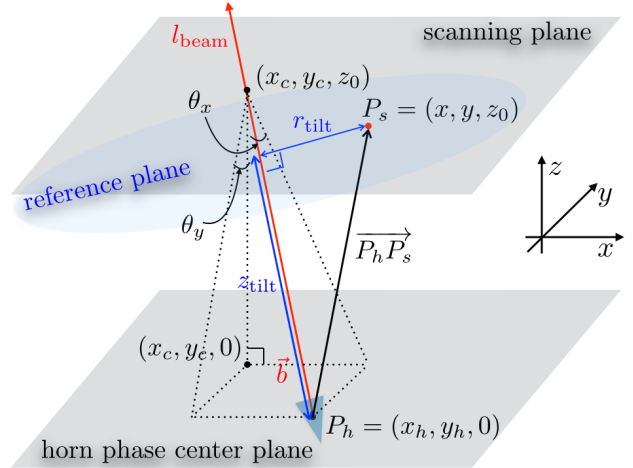

Fig. 4. Schematic of the tilted beam measurement. The feed horn is tilted by $\theta_x$ and $\theta_y$ on the vertical plane in $x$ and $y$-direction. The Gaussian beam radiates from the horn phase center $(x_h, y_h, 0)$ along its axis of beam propagation $l_{beam}$ (red arrow). The beam scanning plane and the horn phase center plane are vertically separated by $z_0$.

We now have the normalized electric field distribution for the tilted beam on the scanning plane

$$E(x, y, z_0) = \left(\frac{2}{\pi w_{tilt}^2}\right)^{0.5} \exp\left(-\frac{r_{tilt}^2}{w_{tilt}^2} - jkz_{tilt} - \frac{j\pi r_{tilt}^2}{\lambda R_{tilt}} + j\phi_{0,tilt}\right),$$

where

$$R_{tilt} = z_{tilt} + \frac{1}{z_{tilt}}\left(\frac{\pi w_0^2}{\lambda}\right)^2,$$

$$w_{tilt} = w_0 \left[1 + \left(\frac{\lambda z_{tilt}}{\pi w_0^2}\right)^2\right]^{0.5},$$

$$\phi_{0,tilt} = \arctan\left(\frac{\lambda z_{tilt}}{\pi w_0^2}\right).$$

The center of the beam that hits the scanning plane is where the equation of line $l_{beam}$ intersects with $z = z_0$ plane and

$$x_c = x_h - z_0 \tan\theta_x,$$
$$y_c = y_h + z_0 \tan\theta_x,$$
$$z = z_0.$$

*B. Beam Fitting*

We fit the beam mapping data to the tilted Gaussian beam profile described in the previous Section IV-A, maximizing the power coupling coefficient between the data and the model electric fields. The power coupling coefficient is an absolute square of the field coupling coefficient

$$K_{ab} = \left|\iint E_a^* E_b dS\right|^2,$$

where $E_a$ and $E_b$ are the electric field distributions of two Gaussian beams. Then, we use Markov chain Monte Carlo (MCMC) sampling with the python package emcee [9].

The corner plot (Fig. 5) is an example of the parameter estimation for one polarization of the 230 GHz beam at $z = 200$ mm. For all scans, the $y$- phase center position is less than 0.5 mm and the $y$-direction tilt angle is less than 0.5 degrees. The best-fit parameters for the receiver tilt angle in the $x$-direction are listed in Table 1 and an example of the model beam is shown in Fig. 3. The 5.74 degrees $x$-direction tilt angle from the design is within 1-sigma uncertainty range of all the measurements.

CONCLUSIONS

We have performed the beam measurement of the SPT VLBI receiver at 230 and 345 GHz. We introduced the functional form of the electric field for the tilted beam. The MCMC fitting of the model to the data gives the estimation of the horn parameters, especially the tilt angle of the feed horn that is critical to the beam coupling between the horn and the optical elements.

ACKNOWLEDGMENT

We thank Edward Tong for useful comments on the measurement technique. This work was supported by NSF awards AST-1207752 and AST-1440254.

TABLE I
THE BEST-FIT PARAMETERS FOR THE $x$-DIRECTION TILT ANGLE

| Tilt angle ($\theta_x$) | | | | | |
|---|---|---|---|---|---|
| 230 GHz | | | 345 GHz | | |
| Pol 0 | $z = 200$ mm | 5.62 (+0.56 / -0.56) | Pol 0 | $z = 195$ mm | 5.80 (+0.64 / -0.64) |
| | $z = 155$ mm | 5.45 (+0.74 / -0.78) | | $z = 150$ mm | 5.69 (+0.88 / -0.89) |
| Pol 1 | $z = 200$ mm | 5.50 (+0.58 / -0.54) | | | |
| | $z = 155$ mm | 5.48 (+0.75 / -0.76) | | | |

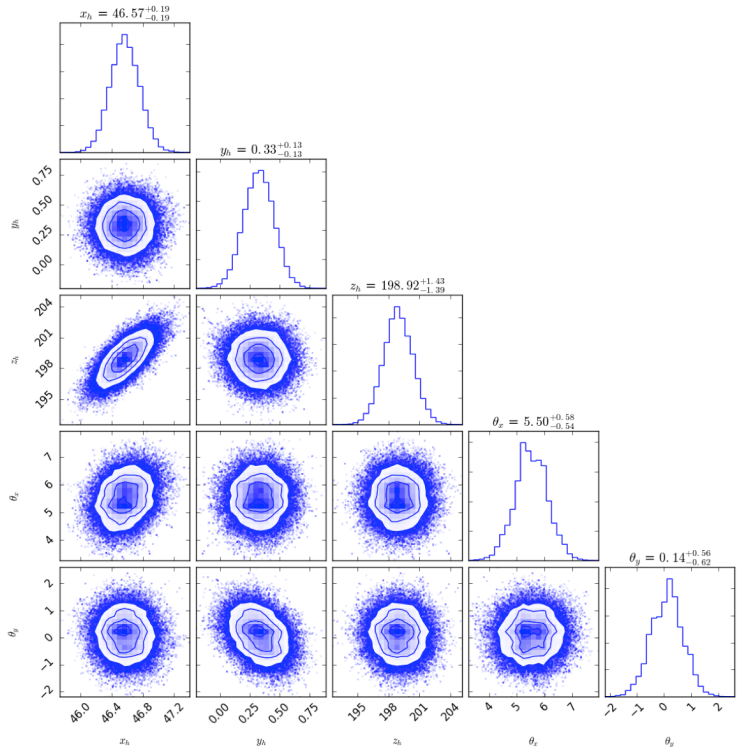

Fig. 5. Parameter estimation result for the 230 GHz measurement at $z = 200$ mm plane using MCMC. The inferred parameters agree well with the design.